# Astronomical Redshifts of Highly Ionized Regions

Peter M. Hansen

ABSTRACT

Astronomical or cosmological redshifts are an observable property of extragalactic objects and have historically been wholly attributed to the recessional velocity of that object. The question of other, or intrinsic, components of the redshift has been highly controversial since it was first proposed. This paper investigates one theoretical source of intrinsic redshift that has been identified. The highly ionized regions of Active Galactic Nuclei (AGN) and Quasi-Stellar Objects (QSO) are, by definition, plasmas. All plasmas have electromagnetic scattering characteristics that could contribute to the observed redshift. To investigate this possibility, one region of a generalized AGN was selected, the so called Broad Line Region (BLR). Even though unresolvable with current instrumentation, physical estimates of this region have been published for years in the astronomical literature. These data, selected and then averaged, are used to construct an overall model that is consistent with the published data to within an order of magnitude. The model is then subjected to a theoretical scattering investigation. The results suggest that intrinsic redshifts, derivable from the characteristics of the ambient plasma, may indeed contribute to the overall observed redshift of these objects.

# 1. INTRODUCTION

The BLR physical data described below were selected in groups by *inspection* for apparent, or approximate, consistency within an order of magnitude. These data are then used to constrain and construct a physical and dynamic model of this region and its electromagnetic scattering characteristics. The approach relies on the fact that the BLR is a hot dense plasma with dielectric properties. Analysis of its dielectric susceptibility provides a basis for examining the scattering or processing of radiation from the central source. This in turn leads to estimates of spectral effects, including an intrinsic redshift. Section 2 focuses on the details of the physical model; Section 3 presents the electrodynamic model; and Section 4 the scattering model. Sections 5 and 6 are, respectively, a brief discussion and the conclusions from this study.

## 1.1 Published Astronomical Data

Figure 1 shows data that have been published in several sources over the period 1985-2010 for the following BLR parameters: (a) Size or radius; and (b) Electron number density. The rectangular regions are the selected data groups which are then arithmetically averaged to obtain generalized physical constraints in the model (shown in the Figure as dashed lines). With increasing radius from the center the numerical averages are: (a) BLR radius: $2.2 \times 10^{13}$ to $1.5 \times 10^{16}$ m; and (b) electron number density: $4.3 \times 10^{17}$ to $4.8 \times 10^{14}$ $m^{-}3$. The temperature and central black hole (BH) mass data are shown in Figure 2 and are taken single valued, again as averages, at respectively: (a) $1.3 \times 10^{4}$ deg K, and (b) $7.5 \times 10^{38}$ kg (or $3.8 \times 10^{8}$ solar masses as shown in the figure). [Note that here and in the rest of this paper all units will be given in Standard International (SI) or *mks* rather than the more familiar *cgs* system of units frequently found in much of the astronomical literature].

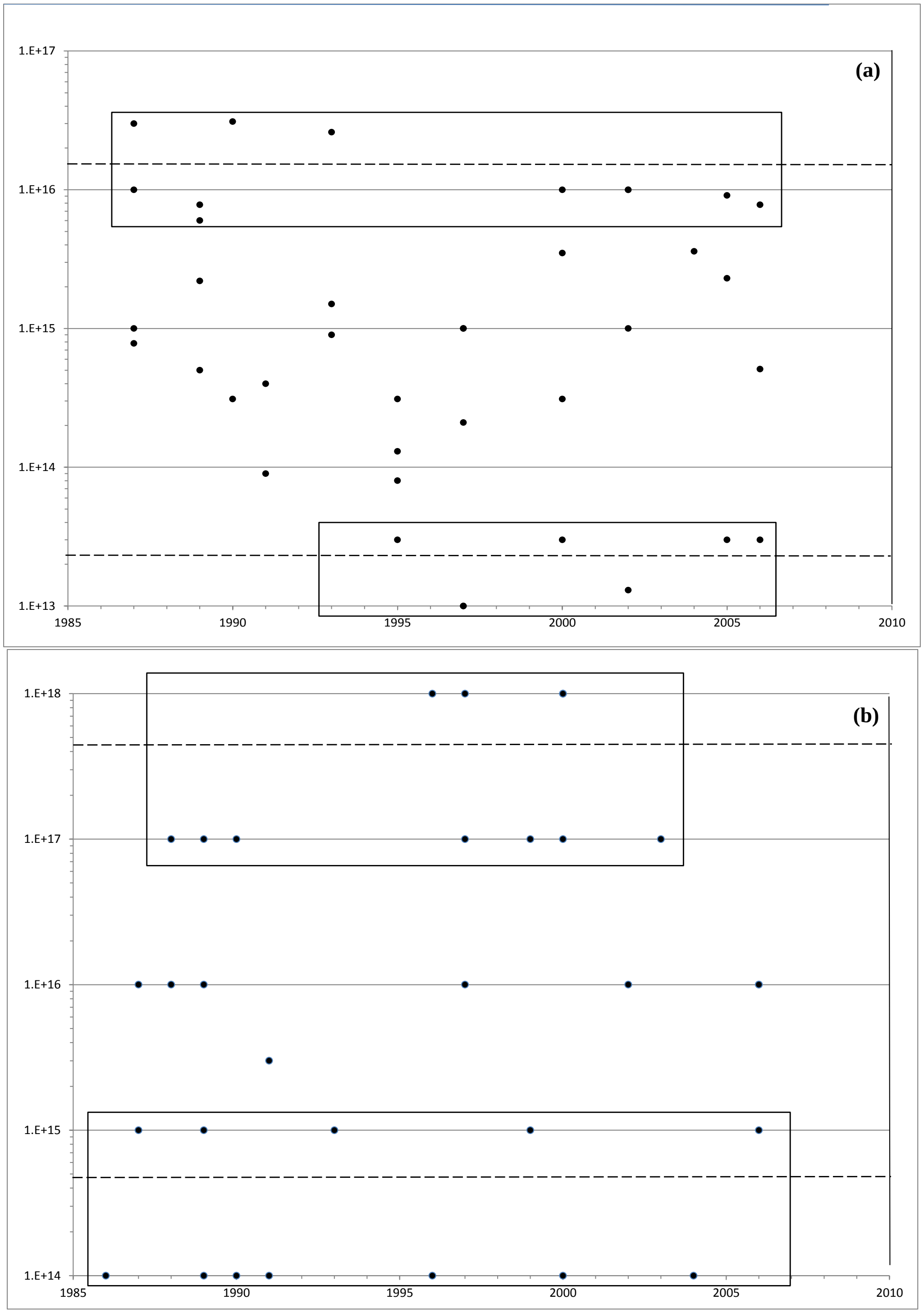

(a) Radius [m]; (b) Electron Number Density [$m^{-3}$]

Figure 1. BLR Size and Electron Number Density

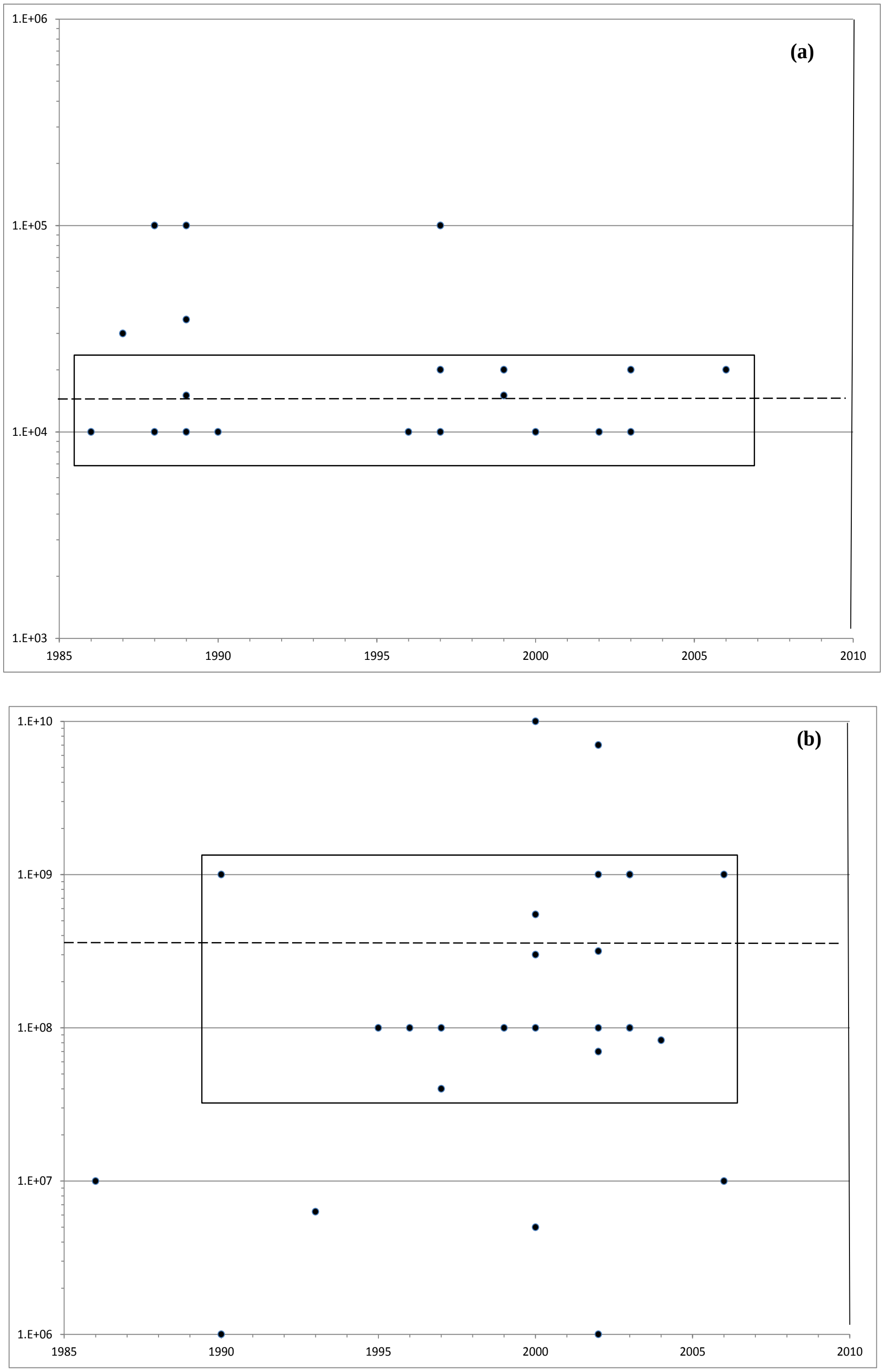

(a) Temperature [degK]; (b) Central Black Hole Mass [solar masses]

Figure 2. BLR Temperature and Central Black Hole Mass

## 1.2. The BLR and its "Clouds"

There appears to be general agreement in the published literature over the years that the BLR is composed of "clouds" of ionized gas (*i.e.* a plasma) that processes and scatters the continuum radiation from the central source into line radiation. Both types of radiation have similar variability, suggesting a surrounding "cloud" structure. For example, from "The BLR is a 'leaky absorber' made of small but dense clouds" (Weedman 1986); to: "The gas in the BLR may be structured in small clumps (hereafter 'clouds')." (Laor et al 2006). The spectral line width indicates Keplerian motion about the central source

Hypotheses on the structure of individual clouds have varied from spherical models (Peterson 1997), to "Clouds are magnetically confined in 'tubes' of circular cross section" (Rees et al 1989); and "the BLR is stratified into multiple emitting zones spread over a wide radial extent" (Baldwin 2003).

The quantitative BLR model developed below attempts to build on these ideas. The physical structure of the clouds is modeled as a set of nested tori, axisymmetric about the central source and containing ionized gas or plasma, confined by its own magnetic field. These clouds are also subject to torsional dynamics arising from poloidal magnetic fields of the adjacent clouds. The size of the clouds is based on a density model and the equilibrium between ionization and recombination similar to the Strömgren criteria. This approach provides for estimates of the number of clouds in the BLR, its total mass, the covering, and the filling factors. In general, these characteristics of the model constructed here are shown below to be within an order of magnitude of published values.

2. THE PHYSICAL MODEL

The BLR electron density model is assumed to be exponential and uses maximum-to-minimum values as a function of radius ($R_1$ to $R_N$ as shown in Figure 1a & 1b discussed above):

1. $n(R) = Ae^{-aR}$

Where:

2. $A = \left\{ [n(R_1) + n(R_N)] \big/ (e^{-aR_1} + e^{-aR_N}) \right\}$

3. $a = \left\{ \ln\left[ n(R_N) \big/ n(R_1) \right] \big/ (R_1 - R_N) \right\}$

The size or radius of the clouds ($r$) as a function of distance from the center ($R$) is defined by the balance [in $s^{-1}m^{-3}$] between the number of ionizations by the central source ($N_\nu$) and the number of recombinations ($N_\alpha$) at the environmental temperature [$T$ in degK from Figure 2a]:

4. $N_\nu(R) = Q\sigma_\nu n(R) \big/ 4\pi R^2$

5. $N_\alpha(R) = n^2(R)\alpha_B(T)$

Where: $Q = Q(T)$ are the photons per sec integrated over the luminosity spectrum of the central source; $\sigma_\nu = 1/[2r\,n(R)]$ is the ionization cross section in the Strömgren criteria (Harwit 2006) where $2r$ is now the cloud diameter rather than the shell thickness of a Strömgren sphere; and $\alpha_B(T)$ is the recombination coefficient summed for all energy levels above ground level, interpolated between 1 x $10^4$ and 2 x $10^4$ degK (Osterbrock & Ferland 2006, Table 2.1). Equating the two numbers ($N_\nu = N_\alpha$) and substituting for the ionization factor: $U(T) = Q(T)/[4\pi R^2 cn(R)]$, the cloud radius is:

$$6.\quad r(R) = U(T)c \big/ 2n(R)\alpha_B(T)$$

The ionization factor used was interpolated at the temperature $T$ from a composite continuum spectrum (Krolik 1999, Figure 10.10). For example, the median radius calculated from equation 6 is within an order of magnitude of the value (~ 3 x $10^8$ m) calculated by Peterson (Peterson 1997).

The BLR was sampled exponentially in radius or distance at 240 points ($R_1$ *to* $R_N$, $N = 240$). Given the radius of the clouds with distance (equation 6), the number of clouds between samples ($R_i$ to $R_{i+1}$) was calculated using an average radii $[(r_{i+1} + r_i)/2]$ for the intermediate cloud diameters ($2r$) and then summing to obtain the total number of clouds ($N_C$):

$$7.\quad N_C = \sum_{i=1}^{N-1}\left[(R_{i+1} - R_i) \big/ (r_{i+1} + r_i)\right]$$

The BLR covering factor ($f_C$) is the total cloud solid angle ($\Omega_i = 4\pi r_i/R_i$) or obscuration as seen from the center. There are two possibilities: if the clouds are in dynamic motion (discussed later) then the maximum $f_c$ would occur if all clouds had unique inclinations (unlikely); the minimum would occur if all inclinations were approximately equal (less likely). These two possibilities are given by:

8. $4\pi \left[{}^{r_i}/_{R_i}\right]_{max} \leq \mathrm{f}_c \leq 4\pi \sum_{i=1}^{N} \left({}^{r_i}/_{R_i}\right)$

The total mass ($M_{BLR}$) follows from the total number of clouds ($N_C$), their toroidal volume and number density (from Figure 1b assuming total ionization: $n_p = n_e = n$) times the proton mass ($m_p >> m_e$):

9. $M_{BLR} = 2\pi^2 m_p \sum_{i=1}^{N} R_i r_i^2 n(R_i)$

Finally, the filling factor ($f_\varepsilon$) is the ratio of the total cloud volume ($V_i = \Sigma_i\ 2\pi^2 R_i\ r_i^2$) to the spherical volume of the BLR:

10. $f_\varepsilon = \left[{}^{\sum_{i=1}^{N} V_i}/_{(4\pi R_N^3/3)}\right] = \left({}^{3\pi}/_{2R_N^3}\right) \sum_{i=1}^{n} R_i r_i^2$

The results of equations 7 – 10 across the BLR radius are shown in Figure 3. Note that published values for these quantities from the references shown in the figure are, for the most part, within an order of magnitude of those calculated in this model.

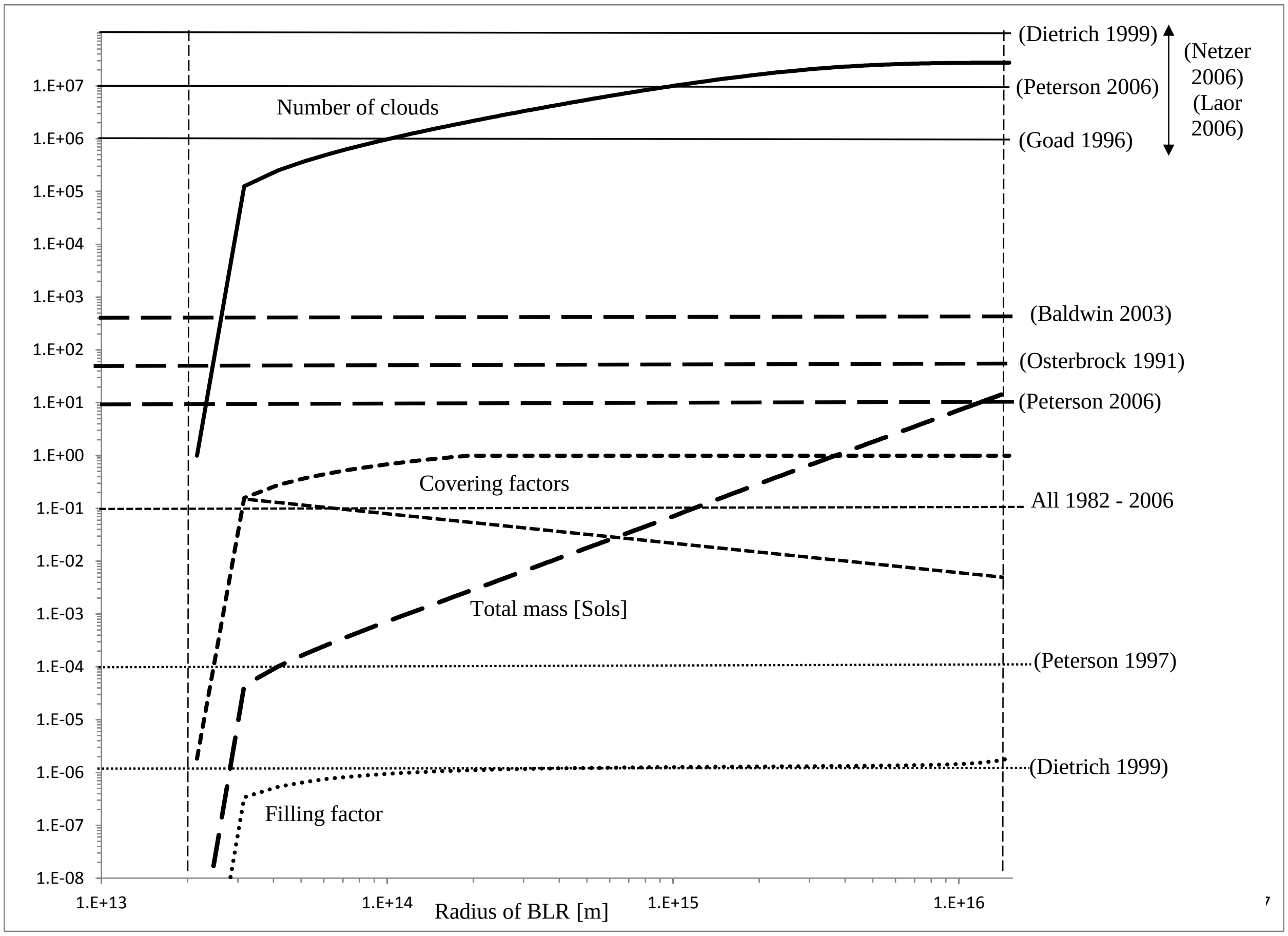

Figure 3. BLR Physical Parameters

## 3. THE ELECTRODYNAMIC MODEL

A diagram of the proposed cloud model and spherical geometry ($R$, $\theta$, $\phi$) of the BLR is shown in Figure 4 for an intermediate (in distance from the center) set of three clouds or tori (the cloud sizes and intercloud distances shown are illustrative only). The Keplerian orbital velocity and current density vectors ($\boldsymbol{V}_\phi$, $\boldsymbol{J}_\phi$) are shown in azimuth ($\phi$). Both types of plasma charges (q = e, and -e) contribute to the current density (Weibel 1959), but due to the difference in mass is primarily electronic. This follows from the particle Hamiltonian for stable Keplerian orbits (assuming no radial acceleration or angular momentum exchange), no gyroscopic oscillation (assuming a weak toroidal or azimuthal galactic magnetic field $B_{\phi} \sim 0$), and no radial or other electric potentials:

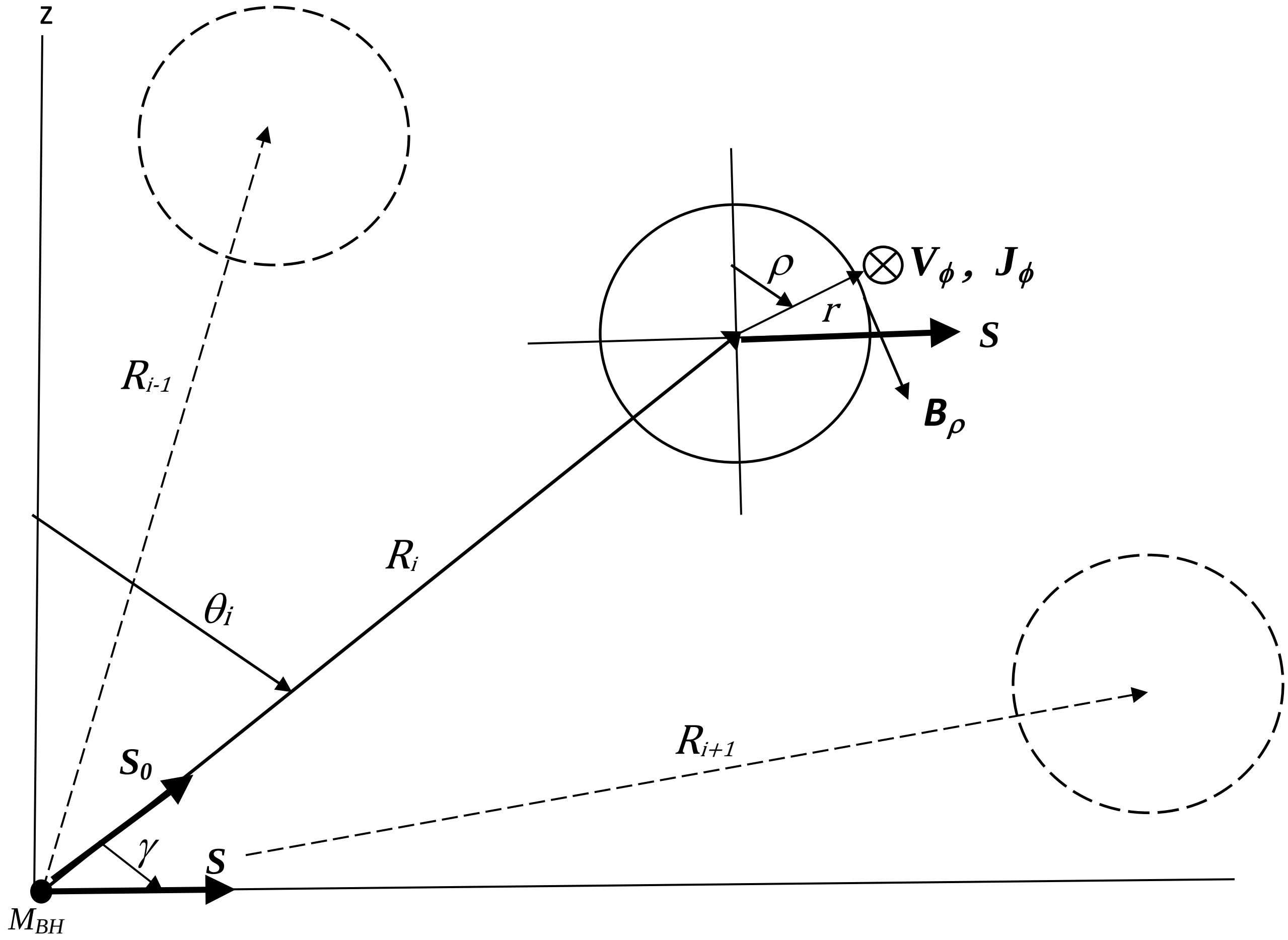

Figure 4. BLR Cloud Structure and Geometry

11. $H = {}^{1}\!/_{2m} \left( {}^{P_{\emptyset}}\!/_{R} - qA_{\emptyset} \right)^{2}$

Where $A_{\phi}$ is the azimuthal vector potential. Since the Keplerian orbital velocity and angular momentum ($P_{\phi}$) is a constant of the motion, then:

12. $\partial H /_{\partial P_{\emptyset}} = 0,\ and\ P_{\emptyset} = mV_{\emptyset}R = qRA_{\emptyset},\ or\ V_{\emptyset} = {}^{qA_{\emptyset}}/_{m}$

The electron current density is:

13. $J_{\emptyset} = nqV_{\emptyset} = A_{\emptyset}q^2\left(\frac{n_e}{m_e} + \frac{n_p}{m_p}\right) \approx {}^{A_{\emptyset}e^2n_e}/_{m_e}$

In a plasma (not at thermal equilibrium) the current density and its magnetic field in the tori produce a classical pinch or a force towards the center (see Figure 4 above) that is assumed here to be the cloud confining mechanism (Weibel 1959; Rees et al 1989). Stability is discussed later.

14. $\boldsymbol{F_r} = \boldsymbol{J}_{\emptyset} \times \boldsymbol{B_\rho} = -F\boldsymbol{e_r}$

Where $\boldsymbol{e_r}$ is the unit radial vector. The azimuthal plasma current ($I_\phi$) at the plasma-vacuum boundary is a toroidal surface current (Webster 2010) resulting from discontinuity in the two magnetic field regions:

15. $\mu_0 I_{\emptyset}\boldsymbol{e}_{\emptyset} = \boldsymbol{e_r} \times (\boldsymbol{B_{\rho>}} - \boldsymbol{B_{\rho<}})$

Where $\boldsymbol{e_\phi}$ is the azimuthal unit vector, and $\boldsymbol{e}_\rho$ is the inner toroidal angle unit vector (see Figure 4). The inner (<) and outer (>) magnetic fields are found, respectively, from magnetohydrodynamic and Ampere's equations (Jackson 1975):

16. $\boldsymbol{\nabla} \times \boldsymbol{B_{\rho<}} = \mu_0\boldsymbol{J}_{\emptyset} = \mu_0 neV_{\emptyset}\boldsymbol{e}_{\emptyset}$

17. $\boldsymbol{B_{\rho<}} = \left({}^{\mu_0 neV_{\emptyset}r_<}/_{2}\right)\boldsymbol{e_\rho}$

18. $\boldsymbol{B}_{\boldsymbol{\rho}>} = \left( \mu_0 I_\emptyset / 2\pi r_> \right) \boldsymbol{e}_{\boldsymbol{\rho}}$, then letting $r_< = r_> = r$ in equation 15:

19. $I_\emptyset = \left[ \pi n e V_\emptyset r^2 / (2\pi r - 1) \right] \to I(n_i, R_i, r_i) \to I_i$ , the azimuthal current.

Since the distance between adjacent clouds can vary, on the order of their distance from the center ($10^{13}$ to $10^{16}$ m) at inclinations above the orbital plane of 30 degrees or more for example, compared to their size or diameter of $10^7$ to $10^8$ m, they are treated below as current "loops" in the overall BLR volume.

The magnetic field from a circular current loop (of radius *R*) has been calculated (Jackson 1975) using Legendre and Associated Legendre functions at a point in spherical coordinates $(r, \theta)$. Note that $r_<$ and $r_>$ are the lesser or greater of *R* or *r*:

20. $B_r = \left( \mu_0 I R / 2r \right) \sum_{n=0}^{\infty} \left[ (-1)^n (2n+1)!!\, r_<^{2n+1} / 2^n n!\, r_>^{2n+2} \right] P_{2n+1}(\cos\theta)$

21.

$B_\theta =$

$$-\left( \mu_0 I R^2 / 4 \right) \sum_{n=0}^{\infty} \left[ (-1)^n (2n+1)!! / 2^n (n+1)! \right] \begin{Bmatrix} -\left( 2n+2 / 2n+1 \right) \left( r^{2n} / R^{3+2n} \right) \\ \left( R^{2n} / r^{3+2n} \right) \end{Bmatrix} P_{2n+1}^1(\cos\theta)$$

The magnetic fields ($B_r$, $B_\theta$) at loop $R_i$ is the superposition of those from $R_{i-1}$ and $R_{i+1}$. Taking just the first two terms in $P_{2n+1}$ and $P^1_{2n+1}$ and letting $\alpha_{i-1} = \theta - \pi/2$, and $\alpha_{i+1} = \pi/2 - \theta$, the magnetic fields at loop $R_i$ are:

22. $$B_r = \left({}^{\mu_0}/_2\right)\left\{\left({}^{I_{i-1}R_{i-1}^2}/_{R_i^3}\right)\left[-\sin\alpha_{i-1} + \left({}^{3R_{i-1}^2}/_{16R_i^2}\right)(3\sin\alpha_{i-1} + 5\sin 3\alpha_{i-1})\right] + \left({}^{I_{i+1}}/_{R_{i+1}}\right)\left[\sin\alpha_{i+1} - \left({}^{3R_i^2}/_{16R_{i+1}^2}\right)(3\sin\alpha_{i+1} - 5\sin 3\alpha_{i+1})\right]\right\}$$

23. $$B_\theta = \left({}^{\mu_0}/_4\right)\left\{\left({}^{I_{i-1}R_{i-1}^2}/_{R_i^3}\right)\left[-\cos\alpha_{i-1} + \left({}^{9R_{i-1}^2\cos\alpha_{i-1}}/_{8R_i^2}\right)\left(5\sin\ {}^2\alpha_{i-1} - 1\right)\right] + \left({}^{I_{i+1}}/_{R_{i+1}}\right)\right\}$$, then:

24. $$B(R_i) = \sqrt{B_r^2(R_{i-1}, I_{i-1}; R_{i+1}, I_{i+1}; R_i) + B_\theta^2(R_{i-1}, I_{i-1}; R_{i+1}, I_{i+1}; R_i)}$$

The adjacent current loops produce a resultant torque on the central loop about the line of intersection in their respective orbital planes (Jackson 1975). The initial conditions and dynamics (torsional oscillations) of these loops are unknown, prohibitively complicated, and not modeled. This dynamic effect may, however, account for the variability seen in the continuum and emission line fluxes attributed to the BLR "clouds" discussed above.

## 4. THE SCATTERING MODEL

The scattering or processing of the continuum radiation into line radiation in the clouds can occur in at least three ways: Compton, stimulated Raman, or stimulated Compton (Krishan 1994). One characteristic of plasma scattering is the spatial correlation of the dielectric susceptibility in the plasma ($\chi$, Bertolotti et al 1976; Mandel & Wolf 1995). The azimuthal Keplerian plasma current normal to the external poloidal magnetic field described above has a dielectric susceptibility (Kulsrud 2005), or *permeability* (Schmidt 1979), similar to that of an azimuthal plasma current driven by an applied electric field, normal to an ambient magnetic field, given by, and estimated here as:

25. $$\chi(R_i) \approx \left[ {n(R_i) m_p} \Big/ {\epsilon_0 B^2(R_i)} \right]$$

Since the magnetic field is a geometric and dynamic quantity, the geometry of Figure 4 used in equations 22 & 23 was generated in a *Monte Carlo* simulation: 100 runs calculating the dielectric susceptibility were made at each of the 240 range samples ($R_i$) across the BLR model described above. Figure 5 shows one set of runs. The darker open circles represent an arithmetic average of the susceptibility $\langle\chi(R_i)\rangle$ over the 100 values calculated at each range sample. Note that the minimum range between clouds is fixed by their number and spacing, hence the maximum magnetic field and minimum susceptibility limit in equation 25 can be seen in Figure 5. The maximum range varies with the simulation geometry producing higher and random values of the susceptibilities. The arrows along the range or radial axis are the locations selected for spatial autocorrelation or autocovariance analysis.

At each radial location ($R_i$) the average susceptibility is calculated across a (Strömgren) cloud diameter $r(R_i)$ from equation 6, or $[R_i - r(R_i)] < r_i < [R_i + r(R_i)]$ in significant decadal steps (1, 2...9 x $10^x$ over the range: $-8 < x < 9$, providing $\langle\chi_i\,(s, R_i)\rangle$, $s \leq 180$ inter cloud range samples.

The magnetic field in equation 25 at each inter cloud range sample was found by interpolation between $B(R_{i-1})$ and $B(R_{i+1})$ using equation 24.

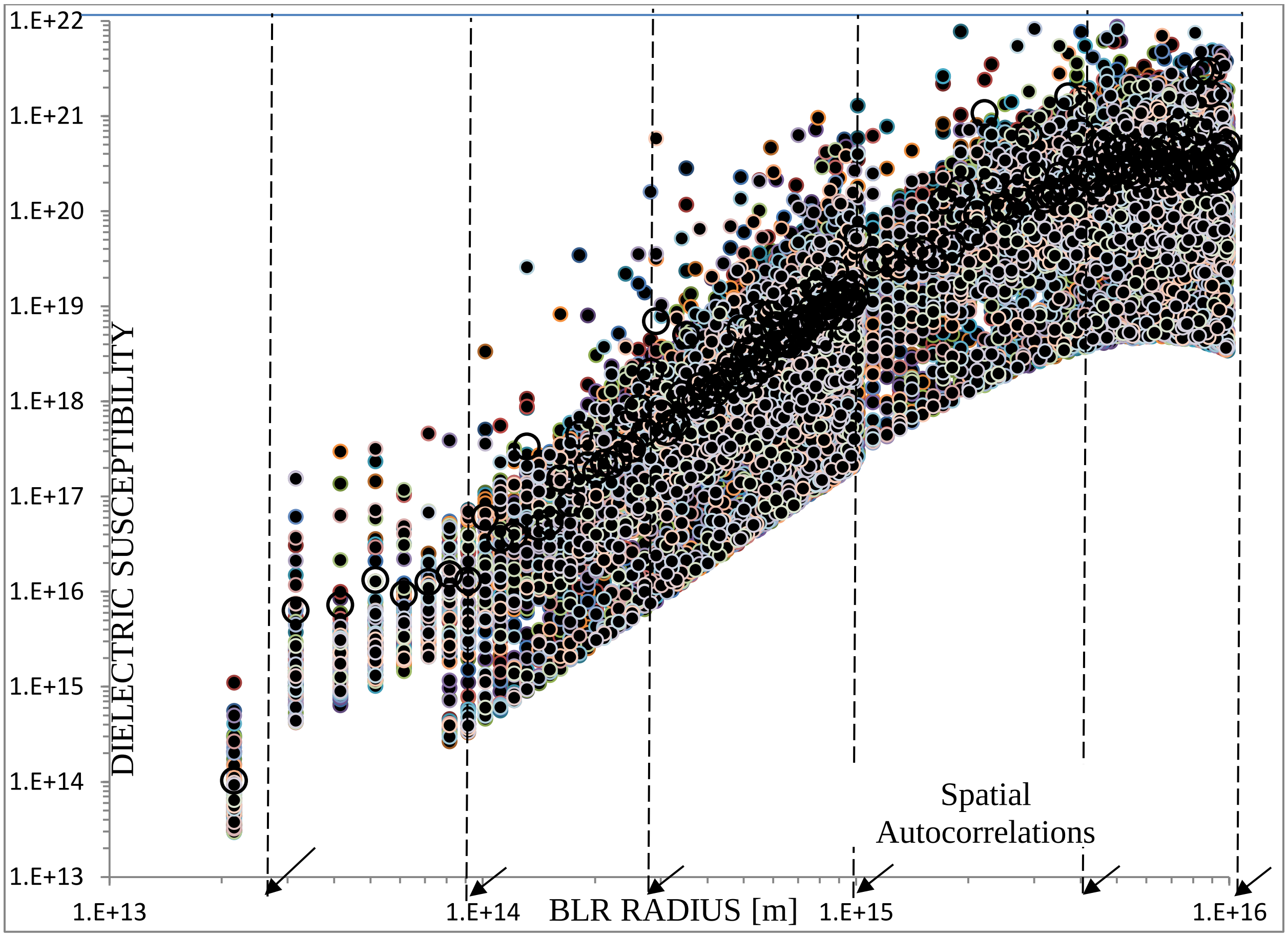

Figure 5. BLR Dielectric Susceptibility

A set of *m* lags (*k*) are used to compute the autocorrelation[1] of the average dielectric susceptibility at each radial location $R_i$ across the cloud extent $r_i$:

26. $C_{\chi(i)}(k) = \left(1/m\right) \sum_{j=1}^{m} < \chi_j^{(i)} >< \chi_{j+k}^{(i)} >$ , $k = 0, 1, 2 \ldots\ldots\ldots m\text{-}1$

These values were then normalized relative to their initial value and then by the sum or integral of the complete set at each radial location:

[1] See Engineering Statistics Handbook, 2012, 1.3.3.1 for using 1/*m* vs. 1/*m*-*k*

27. $\eta_i(k) = \left[\left( {C_{\chi(i)}(k)} \middle/ {C_{\chi(i)}(0)} \right)\right]; \; S_i(k) = \left[ {\eta_i(k)} \middle/ {\sum_{k=0}^{m-1} \eta_i(k)} \right]$

The normalized autocorrelation quantities $S_{(i)}(k)$ are assumed to approximate a Gaussian distribution. Figure 6 shows a set of five runs at selected range locations ($R_i$) and the one-sigma correlation distances in units of $\lambda_0/2\pi$ ("lamda-bar"), where: $\lambda_0 = 2.798 \times 10^{-7}$ m (the Mg II line was selected as an example). Note that the spatial correlation distance of the dielectric susceptibility across a cloud increases as the cloud variability of this quantity decreases with increasing distance from the center ($R_i$, see Figure 5).

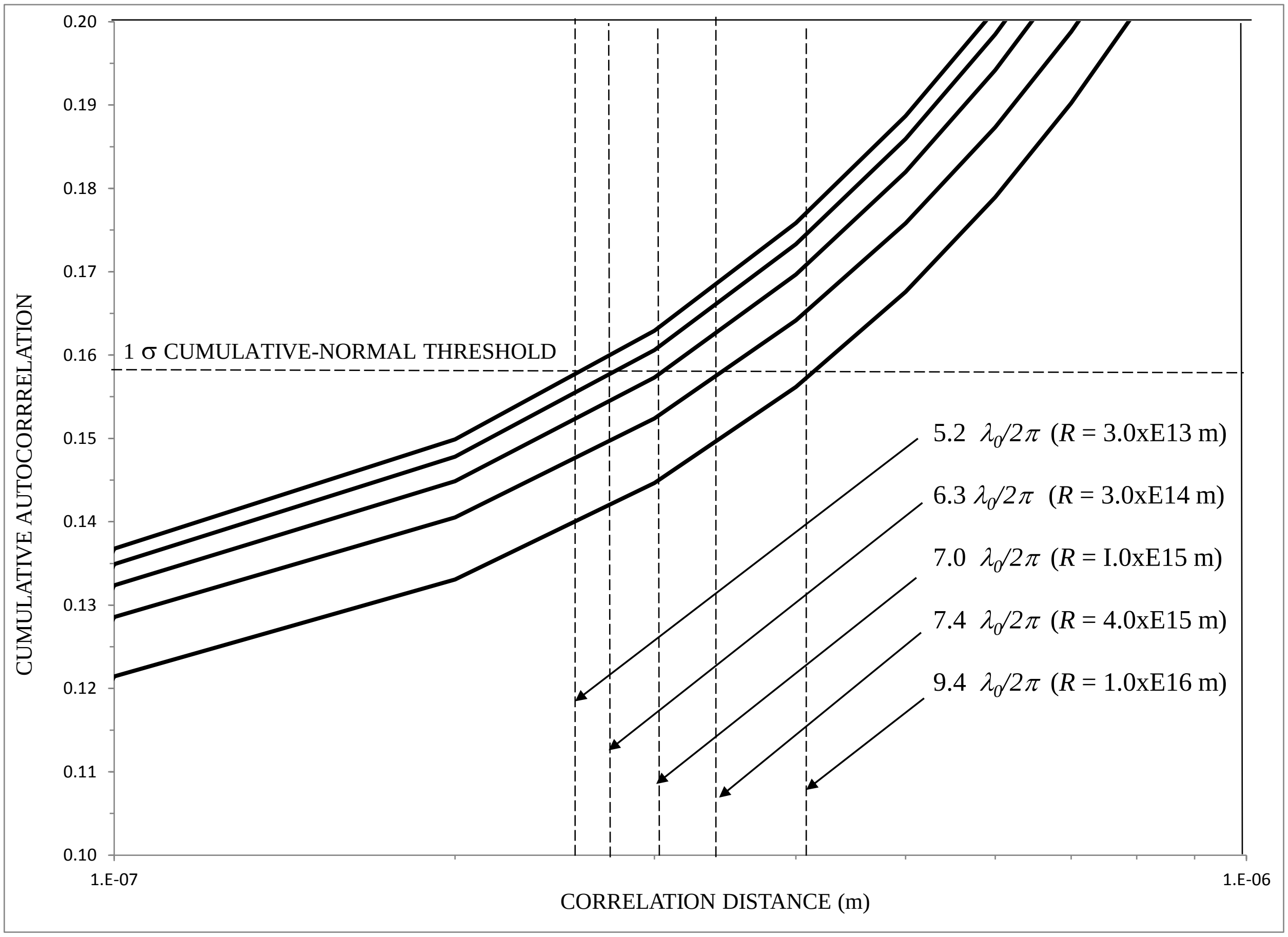

Figure 6. Cumulative-Normal Dielectric Susceptibility Autocorrelations

In 1986 Emil Wolf theoretically predicted that correlations in the dielectric susceptibility at different source points could produce spectral shifts (Wolf 1986). These predictions were later experimentally verified both optically (Gori et al 1988) and even acoustically (Bocko et al 1987). The theory was then extended to scattering media (Wolf et al 1989) and subsequently investigated in great detail (James & Wolf 1990; Wolf & James 1996).

The spectral shift ($z = \Delta\lambda/\lambda_0$) predicted from the scattering of source radiation ($\boldsymbol{S_0}$) at an angle $\gamma$ ($\boldsymbol{S}$, see Figure 4) from spatial correlations in the dielectric susceptibility is (Wolf et al 1989):

28. $$z(\gamma) = 4\left({}^{\Gamma_0}/_{\lambda_0}\right)^2 \left[\left({}^{2\pi\sigma}/_{\lambda_0}\right)^2 \sin^2\left({}^{\gamma}/_{2}\right) - 1\right]$$

Where $\Gamma_0 = 3 \times 10^{-9}$ m is the equivalent Mg II line width (Peterson 1997, Table 1.1); and $\sigma$ is the one-sigma spatial correlation distance across the clouds (Figure 6). The resulting spectral shifts ($z$) from the plasma scattering are shown in Figure 7. Note that at very small scattering angles, less than about 15 degrees, the spectral shift is blue ($z < 0$) but very small ($\approx -10^{-3}$). However, when combined with independent and large recessional redshifts the combination would appear red ($z > 0$). The results of this model, shown in Figure 7, also reproduce previous theoretical predictions (Wolf et al 1989).

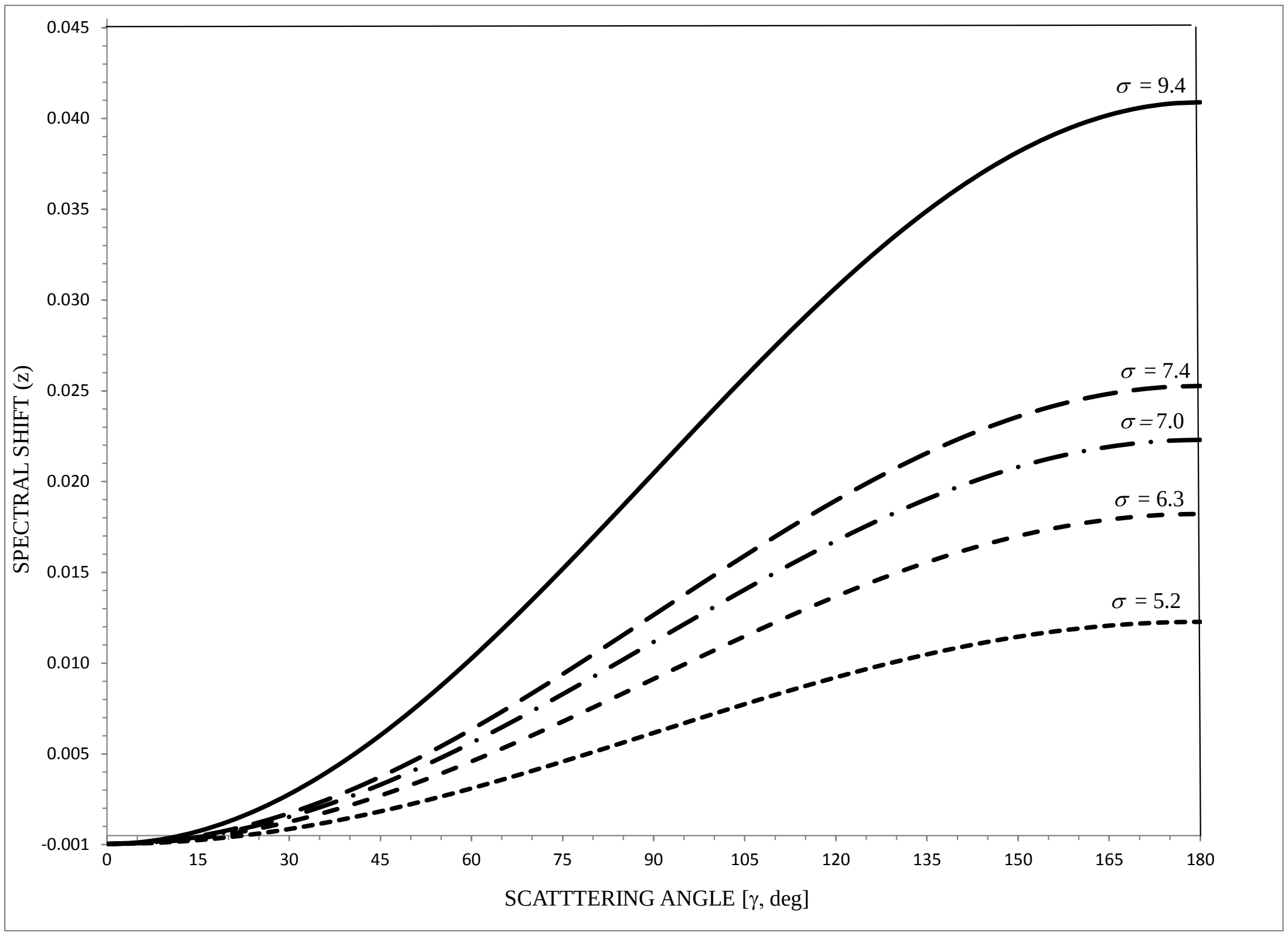

Figure 7. Spectral Shifts from Spatial Correlations in the Dielectric Susceptibility

The observed redshift ($z$) is then a combination of the recessional redshift ($z_1$) and the plasma, or intrinsic, redshift ($z_2$):

29. $z_1 = {(\lambda_1 - \lambda_0)}/{\lambda_0}; \; z_2 = {(\lambda_2 - \lambda_1)}/{\lambda_1}; \; z = (z_1 + 1)(z_2 + 1) - 1$

Figure 8 shows a sequential, by increasing observed redshift ($z$), selection of normal or non-active galaxies (Lang 1980) from the New General Catalog (NGC) with redshifts given in the NASA/IPAC Extragalactic Data Base (NED 2012); also shown are redshifts of AGN, and QSO objects (Cox 2000, Tables 24.3, 24.4). The minimum and maximum plasma redshifts ($z_2$) are shown for comparison (see Figure 7).

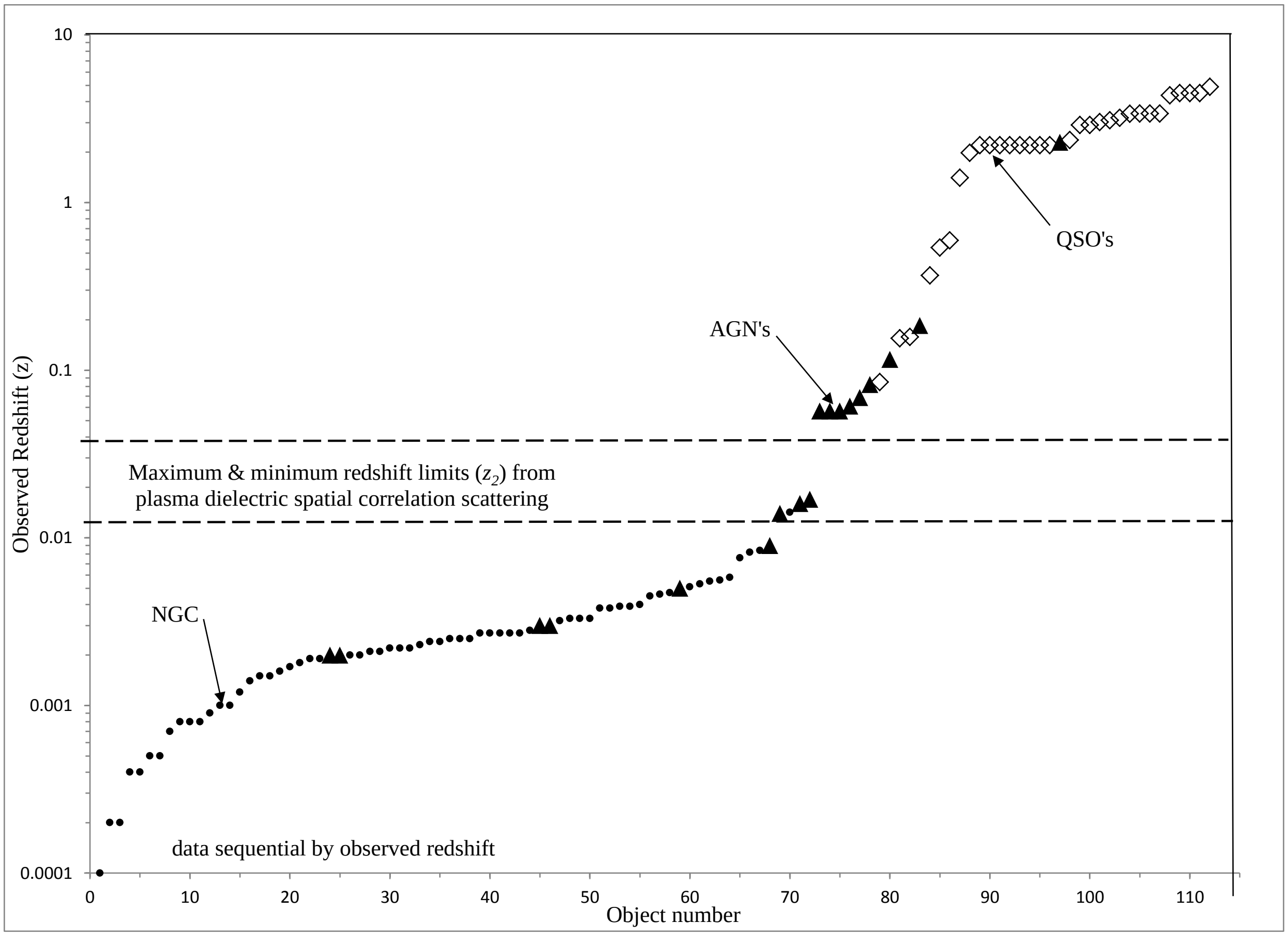

Figure 8. Observed Redshifts

For low redshift objects ($0 < z < z_2$), primarily the normal or non-active (NGC) galaxies, there is not a hot plasma region comparable to the BLR in AGN to produce significant plasma redshifts. The other possibility is that if there is such a region ($z_2 > z > 0$), the recessional redshift could actually be blue ($z_1 < 0$)! In that case:

30. $$z_1 = \left[ {(z - z_2)}\big/{(1 + z_2)} \right] < 0$$

For the higher redshift objects ($z > z_2$) it is of interest to ask what percentage of the observed redshift (Figure 8) is due to recession alone ($z_1/z$). For the maximum plasma redshift: $0.04 = z_2 < z < 1$, the recessional redshift can be 29 – 75% of that observed; for $1 < z < 5$ it increases to 90 – 95%. However, it should be noted that the model developed here was based on AGN estimates (not high $z$ QSO objects) as can be seen in the overlap between the AGN data and the region of plasma redshifts in Figure 8. In this region the recessional redshift percentage is less than 63% and the importance of the plasma redshift is clear. A more accurate model of the plasma regions for the high $z$ objects could reduce the recessional percentages calculated above.

## 5. DISCUSSION

It is clear that the BLR model developed and explored here is, at best, an *average* representation of the physical case fraught with obvious selection bias. Nevertheless, derived results from it are within an order of magnitude of published estimates for AGN's in general. Several important considerations were not addressed: for example, plasma turbulence and the stability of the pinch configuration as the cloud confining mechanism; also, the complicated dynamics of the poloidal magnetic field interactions. These and other effects would be time dependent, but since our observations of the BLR in particular, especially in QSO's (James 1998), are very limited, detailed models have not been developed. Also, the AGN narrow line region (NLR) was not addressed but might be amenable to similar methods.

## 6. CONCLUSION

Up to the present time, all observed redshifts have been interpreted as recessional. It is clear that many important astronomical objects contain hot, dense, plasma regions that, as shown in this study, could provide an additional, or intrinsic, redshift component. These include AGN's, QSO's, and even distant supernovas.